# The Structure, the Dynamics and the Survivability of Social Systems

Ingo Piepers

October 10[th], 2006

*Amsterdam, The Netherlands*
*E-mail: ingopiepers@gmail.com*

**Abstract**

Social systems must fulfil four basic functions to ensure their survival in competitive conditions. Social systems must provide for: (1) energy and other necessities of life, (2) security against external and internal threats, (3) identity and self-development, and (4) consistency and direction. These functions result in four more or less autonomous aspect systems; these aspect systems interact. Between the variables of basic functions and variables of aspect systems, a minimal level of consistency is required to facilitate growth and development, and to ensure the (future) survivability of the social system. Sooner or later, growth, change, and differentiated development result in inconsistencies within/between basic functions and aspect systems. These inconsistencies affect the effectiveness and efficiency of these social systems to provide for the basic functions. Periodically, deliberate or spontaneous corrective adjustments of social systems are required, to correct for these inconsistencies.

## 1. Introduction

In this paper I discuss a framework for the analysis of the structure, the dynamics and the development of social systems. This framework is especially focused on the dynamics and development of the international system[1]. Better understanding of the functioning of the international system is required to improve our ability to anticipate and influence the development of this system.

In this paper I argue that social systems must fulfil four basic functions in order to survive in competitive conditions. In fact, these basic functions work out as 'social organizers' (Boulding, 1978). Furthermore I argue that these functions and corresponding aspect systems interact, and that basic functions and aspect systems need a minimum level of consistency to ensure the effective and efficient functioning of social systems. Next, I explain that as a consequence of growth and change, inconsistencies between aspect systems become unavoidable and sooner or later start to hinder the functioning of social systems: Corrective adjustments become unavoidable to ensure the (future) survivability of the system.

Another fundamental assumption underlying this framework is that growth and development, growth and developmental potential, and the survivability of social systems, are closely related: Population growth unavoidably requires quantitative growth and qualitative development of the four basic functions and corresponding aspect systems of social systems. The potential for growth and development provides a competitive advantage to social systems.

The basic functions of social systems are to a high degree universal. However, different 'underlying' value systems ('cultures') and path dependence can result in fundamentally different dynamics and priorities.

---

[1] Holsti defines an international system as "any collection of independent political entities - tribes, city-states, nations, or empires - that interact with considerable frequency and according to regularized processes. The analyst is concerned with describing the typical characteristic behaviour of these political units toward one another and explaining major changes in these patterns of interaction." (Holsti, 1995)

The outline of this paper is as follows. First I discuss the 'system dynamics' approach - the methodology - I use in this paper to explain the structure and dynamics of basic functions and social systems. Next I define some of the variables of basic functions and corresponding aspect systems in more detail, and I explain the (possible) dynamics of these basic functions and aspect systems from a system dynamics perspective. Then I discuss the (possible) interactions between basic functions, the 'unavoidable' inconsistencies that occur, and corrective adjustments that sooner or later become necessary to ensure the (future) functioning of these systems.

In the next paragraph I present a model for the functioning of corrective adjustments, followed by an explanation of a path dependent dynamic which 'forces' the development of the international system towards a specific configuration.

The purpose of the 'models' I present in this paper is not to solve a particular problem, but to get a better understanding of the structure and dynamics of social systems, and the international system in particular.

## 2. Methodology

This paper is focused on the structure and dynamics of social systems; more specifically on the structure and dynamics of the international system. In this paper I argue that the structure and dynamics of social systems are closely related to the effectiveness, efficiency and survivability of these systems. In this paper I use a systems dynamics approach to 'explain' the workings of social systems and the international system. In this paragraph I explain a system dynamics approach as defined by Sterman (Sterman, 2000).

Sterman argues that the heuristics we use to judge causal relations lead systematically to cognitive maps that ignore feedbacks, multiple interconnections, nonlinearities, time delays, and other elements of dynamic complexity (Sterman, 200, 28). People generally adopt an event-based, open-loop view of causality, ignore feedback processes, fail to appreciate time delays between actions and response and in the reporting of information, do not understand stocks and flows and are insensitive to nonlinearities that may alter the strengths of different feedback loops as a system evolves (Sterman, 2000, 27).

The central assumption of a system dynamics approach is that the behaviour of a system arises from its structure. That structure consists of the feedback loops, stocks and flows, and nonlinearities created by the interaction of the physical and institutional structure of the system (Sterman, 2000, 107). All systems, no matter how complex, consist of networks of positive and negative feedbacks, and all dynamics arise from the interaction of these loops with one another (Sterman, 2000, 13). In other words: the feedback structure of a system determines its dynamics. When multiple loops interact, it is not easy to determine what the dynamics will be. System dynamics emphasizes the multiloop, multistate, nonlinear character of the feedback systems (Sterman, 2000, 21).

Positive (feedback) loops are self-reinforcing. Positive feedback generate growth. It can also create self-reinforcing decline. The loop is self-reinforcing, hence the polarity identifier R.

Negative loops are self-correcting. They counteract and oppose change (Sterman, 2000, 13). Negative loops all describe processes that tend to be self-limiting, processes that seek balance and equilibrium (Sterman, 2000, 12). The B in the centre of a loop denotes a balancing feedback.

The most fundamental modes of behaviour are exponential growth, goal seeking, and oscillation[2]. Exponential growth arises from positive (self-reinforcing) feedbacks (Sterman, 2000, 108).

Negative feedback loops act to bring the state of the system in line with a goal or desired state. They counteract any disturbances that move the state of the system away from the goal. If there is a discrepancy between the desired and actual state, corrective action is initiated to bring the state of the system back in line with the goal (Sterman, 2000, 111- 112).

Oscillation is the third fundamental mode of behaviour observed in dynamic systems. Like goal-seeking behaviour, oscillations are caused by negative feedback loops. In an oscillatory system, the state of the system constantly overshoots its goal or equilibrium state, reverses, and then undershoots, and so on. The overshooting arises from the presence of significant time delays in the negative loop. Time delays - for example in the international system - are caused by the inertia of the international system (as a result of interest of states, the limited ability to assess the actual state of the international system, etc.) and cause corrective actions to continue even after the state of the system reaches its goal, forcing the system to adjust too much, and triggering a new correction in the opposite direction. In the international system the desired goal of the system - accepted by all its actors - is not defined. This characteristic - and the nonlinear interaction of basic feedback structures - contributes to the complex dynamics of the international system.

---

[2] There are other patterns, for example: stasis or equilibrium, in which the state of the system remains constant over time; and random variation (Sterman, 2000, 127).

There are many types of oscillations, including damped oscillations, limit cycles, and chaos. Each variant is caused by a particular feedback structure and set of parameters determining the strengths of the loop and the lengths of the delays (Sterman, 2000, 114).

Often, real systems are nonlinear, meaning that the feedback loops and parameters governing the dynamics vary depending on the state of the system (where the system is operating in state space, the space created by the state variables of the system) (Sterman, 2000, 129).

Causal loop diagrams are simple maps showing the casual links among variables with arrows from a cause to an effect (Sterman, 2000, 102). A causal loop diagram does not show the behaviour (dynamics) of a system, but its structure. Each causal link in a causal loop diagram is assigned a polarity, either positive or negative to indicate how the dependent variable changes, when the independent variable changes. A positive link means that if the cause increases, the effect increases above what it would otherwise have been, and if the cause decreases, the effect decreases below what it would otherwise have been (for a negative link: vice versa).

Link polarities describe the structure of the system. They do not describe the behaviour of the variables. That is, they describe what would happen if there were a change. They do not describe what actually happens.

An increase in a cause variable does not necessarily mean the effect will actually increase. There are two reasons. First, a variable often has more than one input, and second, and more importantly, causal loop diagrams do not distinguish between stocks and flows (Sterman, 2000, 137-140).

Delays are critical in creating dynamics. Delays give systems inertia, can create oscillations, and are often responsible for trade-offs between the short- and long-run effects of policies (Sterman, 2000, 150).

## 3. Basic functions and aspect systems

The definitions of the basic functions of social systems are described in below table. These definitions are based on Boulding[3] (Boulding, 1978) and are focused on states and the international system.

| **Basic Functions of Social Systems** | | |
|---|---|---|
| **Basic function, providing for:** | **Corresponding aspect system** | **Remarks** |
| (1) Energy, necessities of life, and (2) wealth. | Economic system | Provision of energy implies distribution and the availability of a functioning infrastructure |
| (1) Internal and external security and (2) the potential to influence the behaviour of individuals and other (sub) systems. | Threat system | |
| (1) Individual and collective identity and (2) the development of individual and collective identities. | Value system (culture) | |

---

[3] Boulding defines "three major classes of social organizers: the threat relationship, the exchange relationship, and the integrative relationship" (Boulding, 1978, 140). The framework discussed in this paper is based on 'modified' definitions of these social organizers, and a fourth social organizer is added to the framework: The value system.

| Basic Functions of Social Systems (continued) | | |
|---|---|---|
| **Basic function Providing for:** | **Aspect system** | **Remarks** |
| (1) Internal and external consistency, (2) direction for the development of the social system, (3) acceptance of the (political) leadership of the social system, and (4) the possibility to control the environment of the social system. | Integrative system | Control and direction of social systems require a certain degree of predictability and the availability of means and measures to react to changes. Acceptance of the political leadership of a social system is closely related to the (political) objectives of these systems, and the means and measures that can be - and actually are - mobilized to achieve these objectives. |

*Table 1. Variables of basic functions of social systems (states).*

Each aspect system not only has its typical structure, but has its typical - corresponding - 'rules' as well. These rules are closely related to the nature and characteristics of the basic functions.

## 4. Structure and dynamics

In this paragraph I specify the relationships between some of the variables constituting the respective basic functions. The variables and their relationships can change over time: variables are added or fall away, or the polarities of relationships between variables change.

(1) *Energy and other necessities of life*. The provision of energy and other necessities of life are important objectives of the economic system of a state. Depending on the level of sophistication and the priorities of a state, this function is focused on basic needs and/or the provision of wealth. Boulding speaks of an exchange system (Boulding, 1978). In the context of this paper I consider this a somewhat restricted definition. I include energy exploitation and distribution (for instance the physical infra structure of a state) in this function as well.

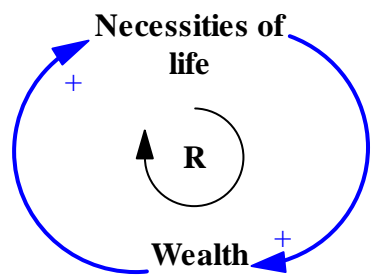

*Figure 1. A causal loop diagram of the economy of a state. These two variables have - in this example - relationships with positive polarities, resulting in a positive feedback loop.*

(2) *Security against external and internal threats*. The threat system of a state must ensure the security of the state and the individuals constituting this social system. External and internal threats can be distinguished. The threat system is not necessarily focused on the improvement of the security of the state in a strict sense. The use of force - or the threat with the use of force - can be aimed at the provision of energy and other necessities of life or at the improvement of the consistency of the (international) system as well.

A threat system can have various (basic) organisational structures. Often integrative structures - the political leadership of states - have control over the use of force. These 'monopolies' can be the result of enforcement (as is the case in dictatorships) and lack acceptance as a result, or can have a (more) 'legal status', as is the case in democracies.

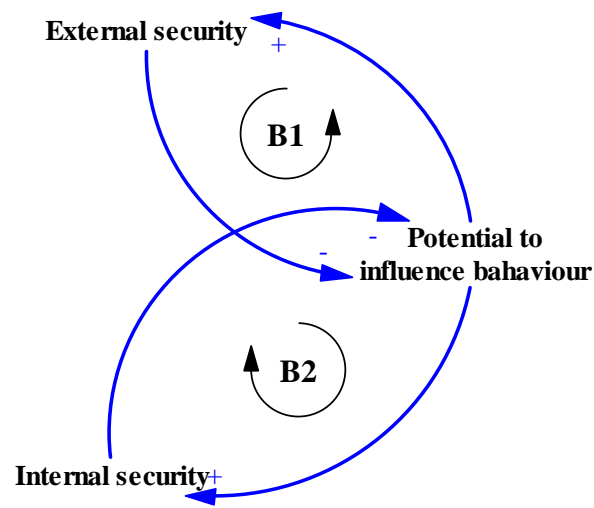

*Figure 2. A causal loop diagram of threat system of a state with three variables and their (possible) relationships: external and internal security, and the potential of integrative structures (actors controlling the threat system(s)) to influence individuals constituting the social system. In this example two negative feedback loops can be identified.*

(3) *Identity and self-development*. Value systems - embedded in the culture of states - provide identities to states and subsystems and individuals constituting these states. Cultures provide stability, consistency and predictability. Value systems are very persistent. However, development of the collective and/or individual identities can become unavoidable in response to changes in the conditions (environment) of social (sub)systems. Sooner or later, lack of adaptation can endanger the survivability of social systems. The 'culture' of a state can be more or less homogeneous; more 'pronounced' individual identities can effect the homogeneity of value systems.

(4) *Consistency and direction*. The provision of consistency and direction are objectives of the integrative systems of social systems (e.g. the political leadership of states). The integrative system is focused on the coordinated fulfilment of the basic functions of social systems and/or the maintenance of the internal and external conditions for the efficient and effective functioning of social systems. Acceptance enhances the effectiveness of integrative systems.

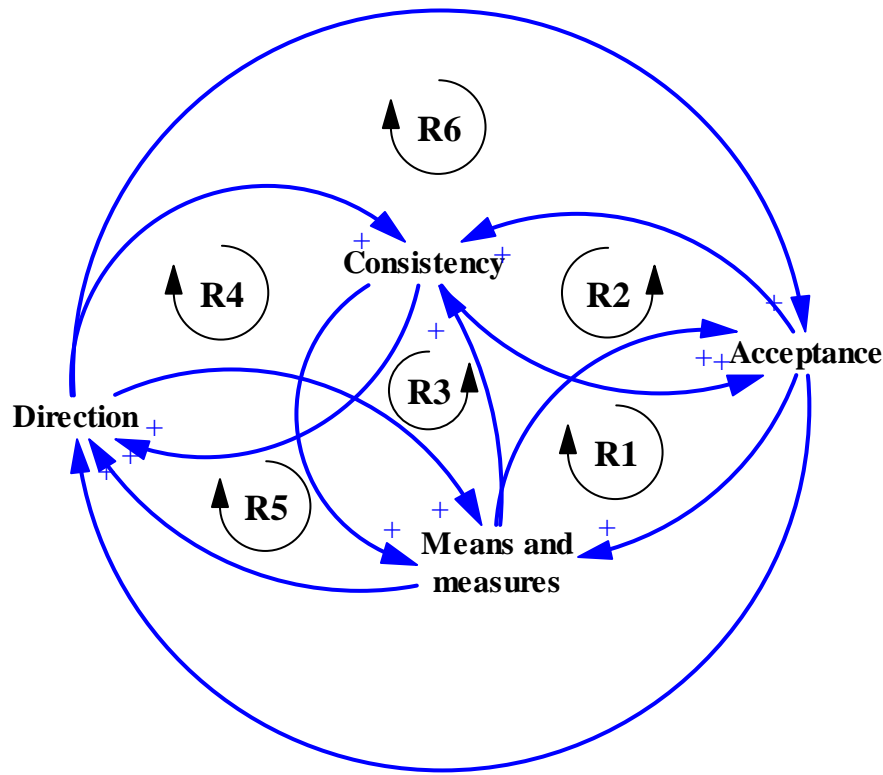

*Figure 3. A causal loop diagram of the integrative system of a state. In this diagram six positive feedback loops can be identified.*

Basic functions and aspect systems of social systems interact and co-evolve; Various dependencies exist between basic functions and aspect systems. Often aspect systems have 'characteristics' of other aspect systems. For example: (1) wealth creation can provide opportunities for the development of collective and individual identities, resulting in a sense of security, and (2) the behaviour of individuals and subsystems can be influenced by economic activities and have an (indirect) effect on the (des)integration of social systems.

## 5. The 'organization' of aspect systems

Basic functions and aspect systems develop their 'own' typical rules, structure and dynamics.
Apart from the type and number of variables of aspect systems and the polarity of relationships between variables, aspect systems differ in connectivity, diversity and degree of clustering. Often aspect systems and their corresponding organizational structures develop their own objectives and priorities, not necessarily to the advantage of other basic functions and aspect systems, or to the advantage of the 'total' system. These autonomous developments contribute to the differentiated development of social systems and result in inconsistencies in these systems.
Social systems develop various mechanisms to deal with inconsistencies and requirements for (unavoidable) change on the one hand, and the maintenance of a certain level of stability and structure on the other hand: democracy is such - and a very effective - 'balancing' mechanism.
The relative importance of the basic functions of social systems varies over time and is dependent on certain system requirements (for instance the requirement for a certain consistency of the system), on environmental conditions, and on (specific) requirements of actors constituting these social systems. These requirements of actors are for instance dependent on the subjective models[4] of these actors and on the relative power position of (dominant) subsystems and coalitions in social systems.

---

[4] In this paper I use the concept of 'subjective models' of actors, constituting social systems. A subjective model of an actor has various components: values and norms (embedded in the value system (culture)), religion,

Aspect systems can be defined as networks. Each aspect system is characterized by its own typical exchanges.
Some examples of networks in states and in the international system:

- *Economic system.* In economic networks goods and services are exchanged Sometimes transformations of goods and services take place during these exchange processes. These networks include the exchange of money/currencies and of information. (Locally) economic networks have typical configurations. These configurations are the outcome of dependencies between actors, power relations, the subjective models of actors, including the incentive structure of the system (North, 1990), and (historic) trade offs.
- *Threat system.* In the network of threat systems, signals are exchanged in the form of implicit or explicit threats. In case these signals fail, force is sometimes used to ensure compliance of other actors. In order to enhance the power position of states, states form coalitions and alliances. These networks have their own typical configurations and characteristics (Kaplan, 1957)[5].
- *Integrative system.* In these networks (inter)actions are coordinated in order to define the desired direction of development, to ensure acceptance, and to implement measures to achieve specific outcomes, etc. These networks can have different configurations, for instance: (con)federative structures or coalitions. The international system is an anarchic system: the integrative structure of the international system is not (yet?) adequately developed to ensure effective integration at a global level[6].
- *Value systems.* In these networks (specific) values are confirmed, rejected or tolerated. Values sometimes change over time. Value systems in the international system - states constituting or representing a specific culture - can have various configurations, for instance depending on the subjective models of (dominant) actors in these value systems, and the existence of a 'core state'. A core state is a dominant actor with the moral authority to influence the behaviour of other actors. This dominant actor in fact represents the value system and has (often) specific 'responsibilities' to defend or safeguard the value system against (external) influence. (Huntington, 1997)[7].

A specific aspect system can - by definition temporarily - dominate a social system. Dominance of a certain aspect system is the result of the interplay between various factors: environmental conditions - for instance a specific threat outside the social system -, the effectiveness of the integrative system (including the functioning of mechanisms which regulate (local) imbalances), the power relations between actors constituting the system, the subjective models of dominant actors, and the level of consistency between the four aspect systems. Dominance of a specific aspect system - often initially functional for the total system - sooner or later requires correction.

---

ethnicity, interests of these actors, and specific experiences which have 'formed' the actor and its 'creation' of reality.

[5] Kaplan differentiates between six configurations (1) the balance of power system, (2) the loose bipolar system, (3) the tight bipolar system, (4) the universal system, (5) the hierarchical system in its directive and non-directive forms, and (6) the unit veto system. (Kaplan, 1957).

[6] It is possible to characterize the international system with this framework. The international system has a more or less global economic network, however wealth is not distributed evenly and the provision of necessities of life is still inadequate. The international system lacks an adequate integrative and threat system: the United Nations lack the acceptance - legitimacy - and means to enforce compliance of international law. The international system does not have a 'generally' accepted value system: discrepancies between values hamper the development of the international system. Inconsistencies between the components of the international system often lead to (local) adjustments - e.g. in the form of conflicts - through which actors try to align the system with their specific interests.

[7] From the perspective presented in this paper, Huntington in fact argues, that after the Cold War discrepancies between value systems dominate the dynamics of the international system. According to the model described in this paper, discrepancies between 'conflicting' value systems need to be addressed - or will be addressed 'spontaneously' - in order to enable growth and development of the 'total' system. Social expansion and the increase of discrepancies are closely related.

As consequence of the dominance of a certain aspect system, the 'rules' of this system and its typical logic will (tend to) dominate the overall dynamics of the total system.

An example: As a result of an external threat, the threat system - including the specific values of this aspect system - can (tend to) dominate the dynamics of a state. As a result, the other aspect systems are often 'pushed' towards the threat system's basic logic and subjective models. Sooner or later, this basic logic sooner will start to hamper the functioning of the other aspect systems; a corrective adjustment then becomes unavoidable.

## 6. Inconsistencies, trade offs and corrective adjustments

The four aspect systems can - at the next level of abstraction - be considered variables of a causal loop diagram. In below figure I assume that the polarities of the relationships between these variables - the respective aspect systems - are positive. This is of course not necessarily the case: it is not difficult to see that the polarity of the relationship between the economy and the threat system can be - or become - negative, for instance as a result of the scarcity of specific resources.

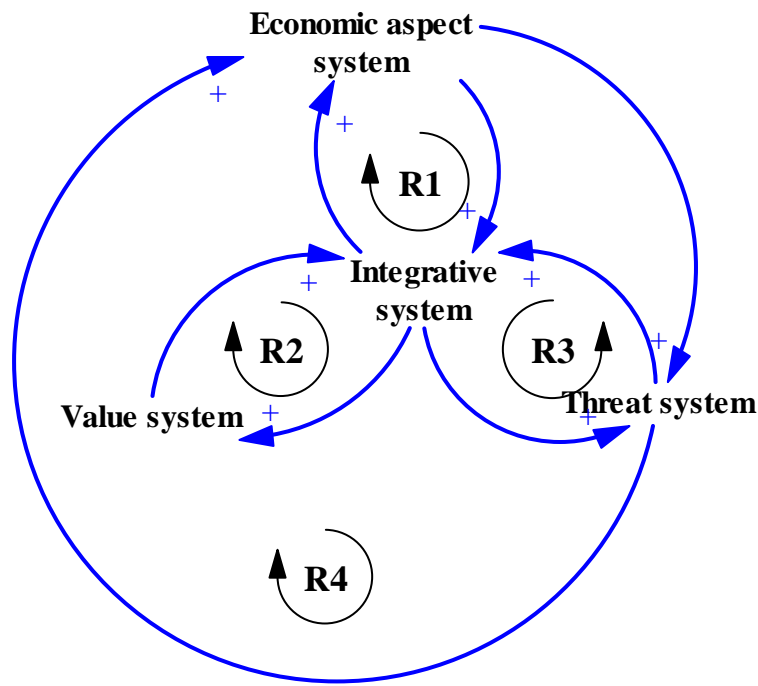

*Figure 4. Example of the relationships between aspect systems constituting a state. Positive feedback mechanisms - growth and change - dominate the dynamics of this particular social system.*

As a result of growth, differentiated development of (components of) social systems, and/or change of the system's conditions, inconsistencies can arise in social systems. Sooner or later these inconsistencies require corrective adjustments - often involving trade offs - to ensure the (future) consistency and survival of these systems.

Gilpin provides examples of inconsistencies and trade offs within states (and within the international system), which - in combination with inertia in these systems - has the potential to cause the decline of states; "a downward spiral" can be the result (Gilpin, 1981, 158).

Gilpin argues that "the national income of a society is distributed into three general sectors: protection, consumption (private and non-military public); productive investment" and "….the historical tendency is for the protection and consumption (private and non-military) shares of national income to increase as a society ages. As a consequence, the efficiency and productivity of the productive sector of the economy on which all else rests will decline". The society enters a downward spiral as a consequence.

As explained, a certain consistency within social systems is required to guarantee the efficient 'operation' of the aspect systems and the provision of basic functions: consistency is a requirement for (future) growth- and development of the system, and its survivability.

Corrective adjustments can be deliberate or be triggered spontaneously. A corrective adjustment is deliberate when it is the result of a decision by the (political) leadership of a state, and when the effects of these corrective adjustments can be controlled effectively. Corrective adjustments are necessary - unavoidably - in the international system as well.

It is problematic that the international system lacks an integrative system which is accepted by all states and ensures positive control of the dynamics of the system. Typically for an anarchic international system, a 'security dilemma' exists or (re)emerges. In an anarchic international system the security of one state (acquired by military power or alliances) 'automatically' implies the insecurity of other states. Such a system tends towards increased hostility. The security dilemma and underlying differences in interests cause complicated nonlinear dynamics.

As a result of the anarchic characteristics of the international system, trade offs and corrective adjustments are often not deliberate and are forced on the system by a more or less autonomous dynamic. In case of such a spontaneous dynamic a positive feedback mechanism dominates the dynamics of the system. In order to better understand the dynamics of the international system before and during such a dynamic (adjustment), it is important to identify and define this positive feedback mechanism and specify the variables of this mechanism and the conditions of the system which 'enable' the 'dominance' of this mechanism (Piepers, 2006).

A spontaneous adjustment can 'emerge' when the political leadership of a social system lacks situational awareness, for instance as a consequence of 'bounded rationality', lacks a sense of direction, lacks the means to influence the dynamics of the social system effectively, or lacks acceptance, limiting the ability to 'align' the system's actors. As a consequence of these 'shortcomings', the adjustment process lacks 'positive control'.

Corrective adjustments - especially spontaneous adjustments - often have more or less disruptive effects on social systems; at least initially. Schumpeter speaks - in the context of innovation in economic systems - of "creative destruction".

Despite these (initially) disruptive effects, social systems often have the potential to reorganize - to readjust the components of the system - and to establish a new form of stability. A minimum level of stability in social systems is required to 'enable' the aspect systems to (again) deliver the basic functions to the system. If a social system lacks the potential to establish such a new balance, the system 'implodes' and loses essential functionalities. This is case with so called failed states.

During a successful readjustment - deliberate or spontaneous - a new optimum is attained, in which the respective aspect systems are better 'balanced' and aligned with each other and the environment of the system. What exactly constitutes an optimum for a specific social system depends for instance on the (environmental) conditions of the system, the internal requirements and the direction of development. However, sooner or later this new optimum will become obsolete and result in new inconsistencies, and - unavoidably - in new corrective adjustments.

In below causal loop diagram the feedback processes - coupling the international system's inconsistencies and resulting inefficiencies to deliberate and spontaneous adjustments processes - are shown. In this model inconsistencies are the result of growth, differentiated development of elements and components (e.g. states) of the international system, and change in (environmental) conditions. The desired - required - state of the system is a state which ensures the fulfilment of the basic functions of states constituting the international system. As explained, it is problematic that in an anarchic system, states often do not share a 'common' - generally accepted - desired goal, and consensus over the means which should be applied to achieve certain outcomes.

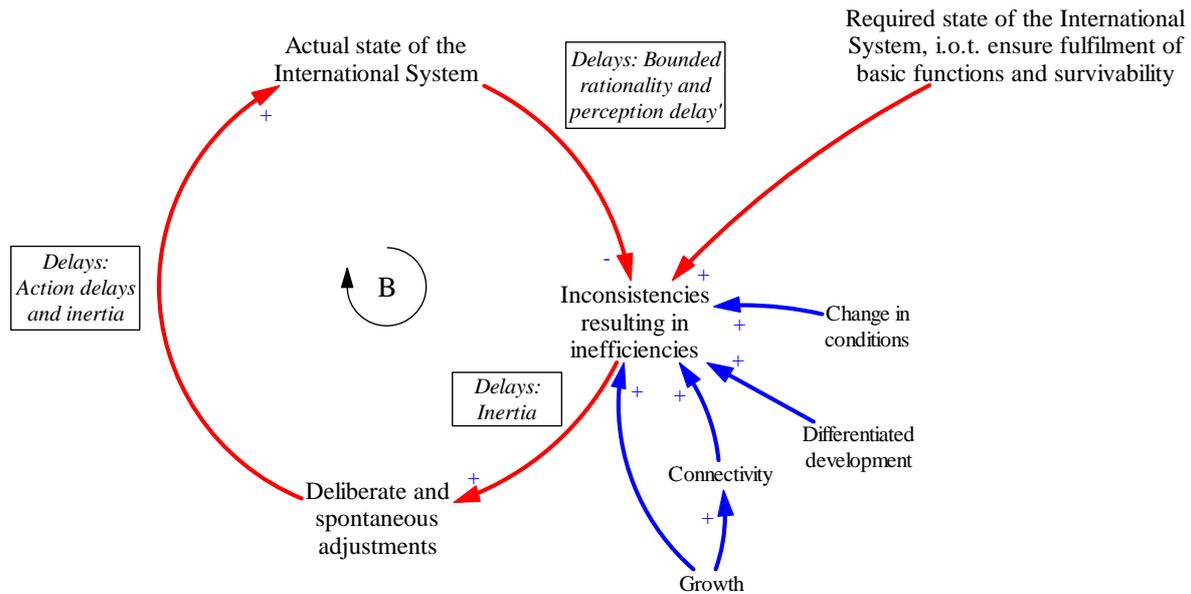

*Figure 5. Goal seeking of the International System: structure and dynamics. The diagram specifies variables - and their relationships - 'constituting' corrective adjustments.*

As a result of the inability of the international system to 'organize' deliberate adjustments, 'corrections - readjustments - of the international system are triggered more or less spontaneously, at least from the perspective of the actors of the international system.

In the paper *"The Dynamics and Development of the International System: A Complexity Science Perspective"*, I argue that (1) large-scale adjustments of the international system - punctuations - require specific (critical) conditions, (2) that these punctuations take place according to a certain logic, (3) that these punctuations are influenced by various network and system effects, and (4) that consecutive punctuations occur and evolve not arbitrarily: It is possible to identify self-organized critical characteristics in the war dynamics of the international system (Piepers, 2006). From 1495 until 1945, during the life-cycle of the European system, four of these punctuations occurred.

During adjustments 'solutions' are selected for the new organization and rules of the international system. In fact, adjustments result in a balancing dynamic; aligning the international system with the interests of the (new) dominant states. As a consequence of the adjustment new growth and developmental potential is created (assuming that the corrective adjustment is successful and does not result in 'functional degradation' of the system).

Dominant powers play an important role in this selection process. Free trade arrangements, the United Nations (with privileged positions for certain (then) dominant states), the International Monetary Fund and the World Bank are examples of the outcomes of selections processes during adjustments. The fact that the outcome of these selection processes to a high degree represent the interests of particular (dominant) actors means that the new international system - for instance its institutions and rules - in fact carries the 'seed' of future inconsistencies, discord and strife. Increasing dissatisfaction with the current international order now is evident and can trigger - if the conditions are favourable - a positive feedback mechanism as discussed in this paragraph.

## 7. Path dependence.

The dynamics of the international system show path dependence. Path dependence is a pattern of behaviour in which the ultimate equilibrium of the system depends on the initial conditions and random shocks and events as the system evolves. In a path dependent system, small, unpredictable events early in the history of the system can decisively determine its ultimate fate and form (Sterman, 2000, 350). Microscopic differences in initial conditions lead to macroscopic differences in outcomes. The eventual end state of a path dependent system depends on the starting point and on small, unpredictable perturbations early in its history. Even when all paths are initially equally attractive, the symmetry is broken by microscopic noise and external perturbations. Positive feedback processes then amplify these small initial differences until they reach macroscopic significance: path dependence arises in systems dominated by positive feedback (Sterman, 2000, 351).

Once a dominant design or standard has emerged, the costs of switching become prohibitive, so the equilibrium is self-enforcing: the system has locked in (Sterman, 2000, 350). Various factors can contribute to the lock in of a particular solution: learning and coordination effects, and adaptive expectations (Arthur, 1988). The path dependent dynamics of the international system are influenced by various system and networks effects as well.

It is important to realize that (different) path dependent dynamics of social systems - sooner or later, especially in case of growth - result in inconsistencies that need to be addressed.

The development of the international system is influenced by advantages of scale, which actors can achieve by the scaling-up of economic activities, and by specialization. Continued economic growth requires the ability of the other aspect systems to 'match' this scaling-up process, otherwise inconsistencies will become unavoidable and growth will be hindered. In case 'balanced' growth of the aspect systems is achieved, a process of social expansion is the result.

Closely related to these processes is the development of the international system towards a condition of greater stability. Two (other) factors - developments - contribute to the increase of stability: the increase of the connectivity of the international system and the increase of thresholds (rules, costs, etc.) for the use of violence against other actors in the system (Piepers, 2006).

These variables - scaling-up of aspect systems, growth, connectivity, etc. - form various positive feedback loops, as is shown in below causal loop diagram.

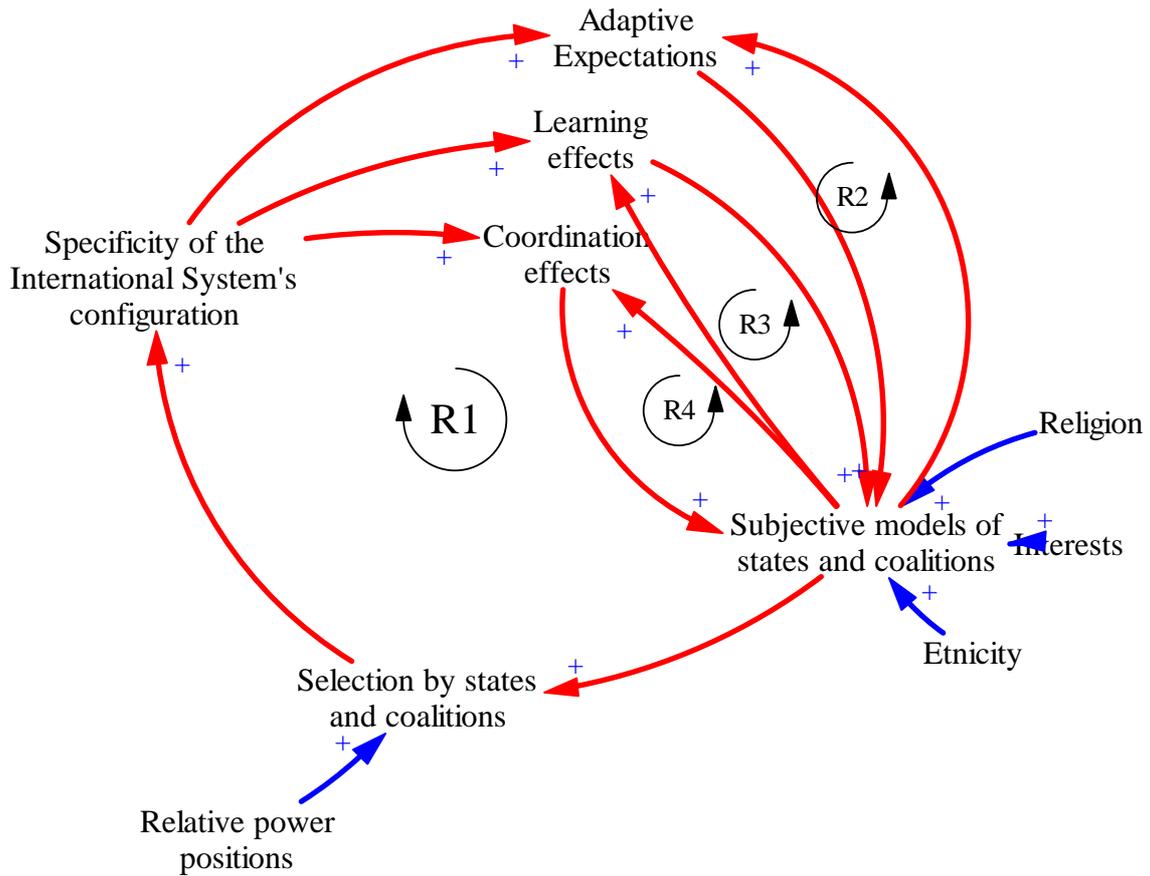

*Figure 6. Path dependence dynamics in the international system.*

At a global level it is possible to identify a certain direction in the development of the basic functions and aspect systems of the international system.

| Aspect systems of the international system | Direction of development |
| --- | --- |
| Economic system | Towards capitalist models. |
| Threat system | Towards the ability to 'counter' large- and small-scale threats with a (potentially) global impact, the ability to fight wars 'amongst the people'. |
| Value system | Towards increased empowerment of individuals, organizations and other subsystems, and towards more accentuated collective and individual identities. |
| Integrative system | Towards a dynamic with the simultaneous (1) expansion of specific structures to enable (global) control and coordination, e.g. in order to coordinate economic activities, achieve economies of scale, and/or to counter (global) threats, and (2) fragmentation of structures to 'secure' and promote 'local' identities. The fragmentation of the international system becomes obvious when the increase of the number of states is taken into account, especially after the Second World War. |

*Table 2. Development of basic functions and aspect systems.*

## 8. Future research.

This is a first effort to develop a framework - based on a system dynamics approach - which 'maps' the structure of social systems, describes the possible relationships between the variables of these systems, and links the structure of the system with its dynamics. Furthermore, I have tried to develop two models, respectively describing the process of corrective adjustments of the international system, and the dynamics which channel the development of social systems.

In another paper, I will test the hypothesis - based on this theoretical framework - that a causal relationship exists between population growth, the life-span of international systems (the period between two consecutive corrective adjustments of the international system), and the optimization of basic functions.